\documentclass[conference]{IEEEtran}
\IEEEoverridecommandlockouts
\usepackage{cite}
\usepackage{amsmath,amssymb,amsfonts}
\usepackage{algorithmic}
\usepackage{graphicx}
\usepackage{textcomp}
\usepackage{xcolor}
\usepackage{float}
\usepackage{threeparttable} 
\usepackage{soul}
\usepackage{hyperref}
\usepackage{multirow}
\usepackage{etoolbox}

\def\BibTeX{{\rm B\kern-.05em{\sc i\kern-.025em b}\kern-.08em
    T\kern-.1667em\lower.7ex\hbox{E}\kern-.125emX}}

\begin{document} 
\title{A Low-Dissipation and Scalable GEMM Accelerator with Silicon Nitride Photonics}

\makeatletter
\newcommand{\linebreakand}{%
  \end{@IEEEauthorhalign}
  \hfill\mbox{}\par
  \mbox{}\hfill\begin{@IEEEauthorhalign}
}
\makeatother

% \author{\IEEEauthorblockN{Venkata Sai Praneeth Karempudi, Sairam Sri Vatsavai,  Oluwaseun Adewunmi Alo, Ishan Thakkar, Oluwaseun Adewunmi Alo, Jeffrey Todd Hastings, and Justin Scott Woods}
% \IEEEauthorblockA{\textit{Department of Electrical and Computer Engineering, University of Kentucky, Lexington, USA } \\
% kvspraneeth@uky.edu, ssr226@uky.edu, igthakkar@uky.edu,  seun.alo@uky.edu,todd.hastings@uky.edu and  }}

\author{\IEEEauthorblockN{Venkata Sai Praneeth Karempudi}
\IEEEauthorblockA{\textit{Department of ECE} \\
\textit{University of Kentucky}\\
Lexington, Kentucky, USA \\
kvspraneeth@uky.edu}
\and
\IEEEauthorblockN{Sairam Sri Vatsavai}
\IEEEauthorblockA{\textit{Department of ECE} \\
\textit{University of Kentucky}\\
Lexington, Kentucky, USA \\
Sairam\_Srivatsavai@uky.edu}
\and
\IEEEauthorblockN{Ishan Thakkar}
\IEEEauthorblockA{\textit{Department of ECE} \\
\textit{University of Kentucky}\\
Lexington, Kentucky, USA \\
igthakkar@uky.edu}
\and
\linebreakand
\IEEEauthorblockN{Oluwaseun Adewunmi Alo}
\IEEEauthorblockA{\textit{Department of ECE} \\
\textit{University of Kentucky}\\
Lexington, Kentucky, USA \\
seun.alo@uky.edu}
\and

\IEEEauthorblockN{Jeffrey Todd Hastings}
\IEEEauthorblockA{\textit{Department of ECE} \\
\textit{University of Kentucky}\\
Lexington, Kentucky, USA \\
todd.hastings@uky.edu}
\and
\IEEEauthorblockN{Justin Scott Woods}
\IEEEauthorblockA{
\textit{Argonne National Laboratory}\\
Chicago, Illinois, USA \\
jwoods@anl.gov}
}

\makeatletter
\patchcmd{\@maketitle}
  {\addvspace{0.5\baselineskip}\egroup}
  {\addvspace{-1.8\baselineskip}\egroup}
  {}
  {}
\makeatother

\maketitle

\begin{abstract}
Over the past few years, several microring resonator (MRR)-based analog photonic architectures have been proposed to accelerate general matrix-matrix multiplications (GEMMs), which are found in abundance in deep learning workloads. These architectures have dramatically grown in popularity because they offer exceptional throughput and energy efficiency compared to their electronic counterparts. However, such architectures, due to their traditional realization based on the silicon-on-insulator (SOI) material platform, face two shortcomings. First, the high-index contrast of the SOI platform incurs high scattering losses, which mandates the provisioning of high optical input power. Second, SOI waveguides are susceptible to two-photon absorption (TPA), which can incur substantial optical signal losses at moderate-to-high signal fan-in. These shortcomings have severely detrimental effects on the achievable parallelism, throughput, and energy efficiency of SOI MRR-based GEMM accelerators. To address these shortcomings, we present a novel Silicon Nitride (SiN)-Based Photonic GEMM Accelerator called SiNPhAR. SiNPhAR architecture employs SiN-based active and passive devices to implement analog GEMM functions. Since the SiN material exhibits lower index contrast and no TPA, the optical signal losses in our SiNPhAR architecture are very low. This advantage significantly enhances the achievable processing parallelism, throughput, and energy efficiency of SiNPhAR architecture, compared to SOI-based photonic GEMM accelerators from prior work. We quantify and compare these benefits of SiNPhAR architecture via our cross-layer evaluation for a benchmark workload comprising four modern deep neural network models. From the system-level performance analysis, SiNPhAR demonstrates at least 1.7$\times$ better throughput (frames-per-second (FPS)) while consuming at least 2.8$\times$ better energy efficiency (FPS/W) than prior SOI-based GEMM accelerators.
\end{abstract}

\section{Introduction}

Deep Neural Networks (DNNs) have revolutionized the implementation of various artificial intelligence tasks, such as image recognition, language translation, autonomous driving \cite{dnnapplications1,dnnapplications2}, due to their high inference accuracy. However, DNNs are computationally intensive, due to inherently abundant linear computations such as general matrix-matrix multiplications (GEMM), which are at the core of DNN operations \cite{demirkiran2021electro}. This computational intensity of processing the GEMM functions of DNNs is on a rapid rise owing to the ongoing rapid evolution of DNN models. This has pushed for highly customized hardware GEMM accelerators \cite{Baischer2021}. Among GEMM accelerators demonstrated in the literature, silicon-photonic accelerators have shown great promise to provide unparalleled parallelism, ultra-low latency, and high energy efficiency \cite{holylight,squeezelight,deapcnn,karen2020proceeding,amm,sunny2021crosslight}. In particular, Microring Resonator (MRR)-enabled silicon-photonic GEMM accelerators have shown disruptive performance and energy efficiencies, due to the compact footprint of MRRs, low dynamic power consumption of MRRs, and the ability of MRRs to support a massive fan-in of optical signals through dense-wavelength-division multiplexing (DWDM). These advantages have rendered up to 1000$\times$ more processing throughput and up to 100$\times$ better energy efficiency to MRR-enabled silicon-photonic GEMM accelerators than their electronic counterparts \cite{deapcnn,acceleratorssurvey}.

%Typically, a silicon-photonic DNN accelerator comprises multiple dot product units that operate concurrently, enabling parallel execution of multiple dot product operations created by unrolling the input GEMM operations. Several GEMM accelerators realized based on the traditional Silicon-on-insulator(SiO$_{2}$) (SOI) photonic platform have been demonstrated in prior works based on various photonic devices, such as the Mach Zehnder Interferometer (MZI) \cite{sunny2021survey} and the Microring Resonator (MRR) \cite{sunny2021survey}. However, the use of Mach-Zehnder Interferometers (MZIs) introduces a significant area overhead, making MZI-enabled DNN accelerators impractical for scaling in large-scale neural networks \cite{sunny2021survey, cheng2020silicon}.

However, the state-of-the-art MRR-enabled GEMM accelerators that are realized using the traditional silicon-on-insulator (SOI) material platform face two shortcomings. First, the high refractive index contrast between the silicon core (Si) and cladding (SiO$_{2}$) of an SOI waveguide leads to an enhanced interaction of the guided optical mode with the rough sidewalls of the waveguide. This introduces high scattering losses in the SOI waveguides \cite{baets2016silicon}. Second, the presence of Two-Photon Absorption (TPA) in silicon has detrimental effects on SOI devices, particularly waveguides. These effects lead to substantial absorption losses in SOI waveguides, particularly when a moderate-to-high number of multiplexed optical signals are propagating inside an SOI channel waveguide. To counter these losses, a higher input optical power becomes necessary. However, this increased input optical power whittles down a significant part of the optical power budget, significantly hampering the achievable spatial parallelism, throughput, and energy efficiency of SOI-based photonic GEMM accelerators.

To address these shortcomings, we present a novel Silicon Nitride (SiN)-on-SiO$_{2}$-based photonic GEMM accelerator named SiNPhAR. Our SiNPhAR accelerator integrates Indium Tin Oxide (ITO)-based SiN-on-SiO$_{2}$ MRR modulators (MRMs) within its input and weight banks, that are coupled to SiN-on-SiO$_{2}$ waveguides. These MRMs perform high-speed electro-optical encoding of electrical inputs and weights onto optical signals. These input and weight banks seamlessly integrate with our invented balanced photo-charge accumulator (BPCA) to perform dot product operations of a large size. Unlike the SOI platform, the SiN-on-SiO$_{2}$ platform has a lower refractive index contrast between the core (SiN) and cladding (SiO$_{2}$) materials. This enables the design of ultra-low loss ($<$0.5 dB/cm \cite{ophir2010demonstration,baets2016silicon}) photonic waveguides. Additionally, the absence of free-carriers in the SiN material eliminates the possibility of TPA \cite{borghi2021modeling,baets2016silicon}. This characteristic enables SiN-on-SiO$_{2}$ photonic waveguides to support a higher count of multiplexed optical signals (higher fan-in) without incurring excess absorption or scattering losses. Reduced optical losses empower our SiNPhAR accelerator to achieve superior spatial parallelism, enhanced throughput, and energy efficiency compared to prior SOI-based photonic GEMM accelerators.

Our key contributions in this paper are summarized below:

\begin{itemize}

\item We present the structure, operation, and characterization results of our designed ITO-based SiN-on-SiO$_{2}$ MRM;

\item  We illustrate the use of our ITO-based SiN-on-SiO$_{2}$ MRMs as input and weighting elements, enabling massively parallel multiplication operations; 

\item We design an accelerator architecture called SiNPhAR, which is based on the SiN-on-SiO$_{2}$ platform, and evaluate its achievable spatial parallelism;

\item We compare the throughput and energy efficiency results of our SiNPhAR architecture with an SOI-based MRR-enabled GEMM accelerator from prior works.

\end{itemize}

\section{Preliminaries}
\subsection{Background on SOI-Based Photonic GEMM Accelerators}
%Write about MZI based coherent and MRR based incoherent accelerator architectures.
Among the SOI-based photonic GEMM accelerators showcased in the literature, we focus on the MRR-enabled SOI-based incoherent GEMM accelerators\cite{al2022scaling,deapcnn,sunny2021crosslight,shiflett2020pixel,cases2022,vatsavai2023sconna}. These accelerators mainly employ multiple analog tensor processing cores (TPCs) that operate in parallel, in which each TPC is utilized to perform a dot product operation. Typically, each TPC is made up of five essential blocks \cite{cases2022} (see Fig. \ref{SOI_MWA} for an example TPC organization with five blocks): (i) a laser block that employs \textit{N} laser diodes (LDs) to generate \textit{N} optical wavelength channels; (ii) an aggregation block that aggregates the optical wavelength channels generated by LDs into a single photonic waveguide through DWDM technique by employing a \textit{N$\times$1} multiplexer, and then splits the optical power of each of these wavelength channels equally into \textit{M} separate waveguides by using a \textit{1$\times$M} splitter; (iii) a modulation block that consists of \textit{M} banks of MRMs spread across \textit{M} dot product elements (DPEs), with each DPE employing one MRM bank; (iv) a weighting block that consists of another \textit{M} banks of MRRs spread across the \textit{M} DPEs, with each DPE employing one MRR bank; and (v) a summation block that comprises of a total of \textit{M} summation elements (SEs), with each SE corresponding to a DPE and employing two photodiodes in a balanced configuration, commonly referred to as balanced photodiode (BPD) configuration, connected to a transimpedance amplifier (TIA) and an analog-to-digital converter (ADC). Typically, the laser block and SE block are placed at the two ends of the TPC, whereas the aggregation, modulation, and weighting blocks are placed in between them. Furthermore, based on the positioning of these intermediate blocks, the MRR-based TPC organizations demonstrated in the prior works can be classified into three categories namely \textit{Aggregate, Modulate, Weight (AMW)} TPC, \textit{Modulate, Aggregate, Weight (MAW)} TPC, and \textit{Modulate, Weight, Aggregate (MWA)} TPC. In the AMW TPC, the aggregation block is positioned first, followed by the modulation and the weighting blocks. On the other hand, in the MAW TPC organization, the modulation block is positioned first, followed by the aggregation and the weighting blocks. For additional details on the AMW and MAW TPCs, we direct the reader to \cite{cases2022}. 

\begin{figure}[h!]
    \centering
    \includegraphics[width = \linewidth]{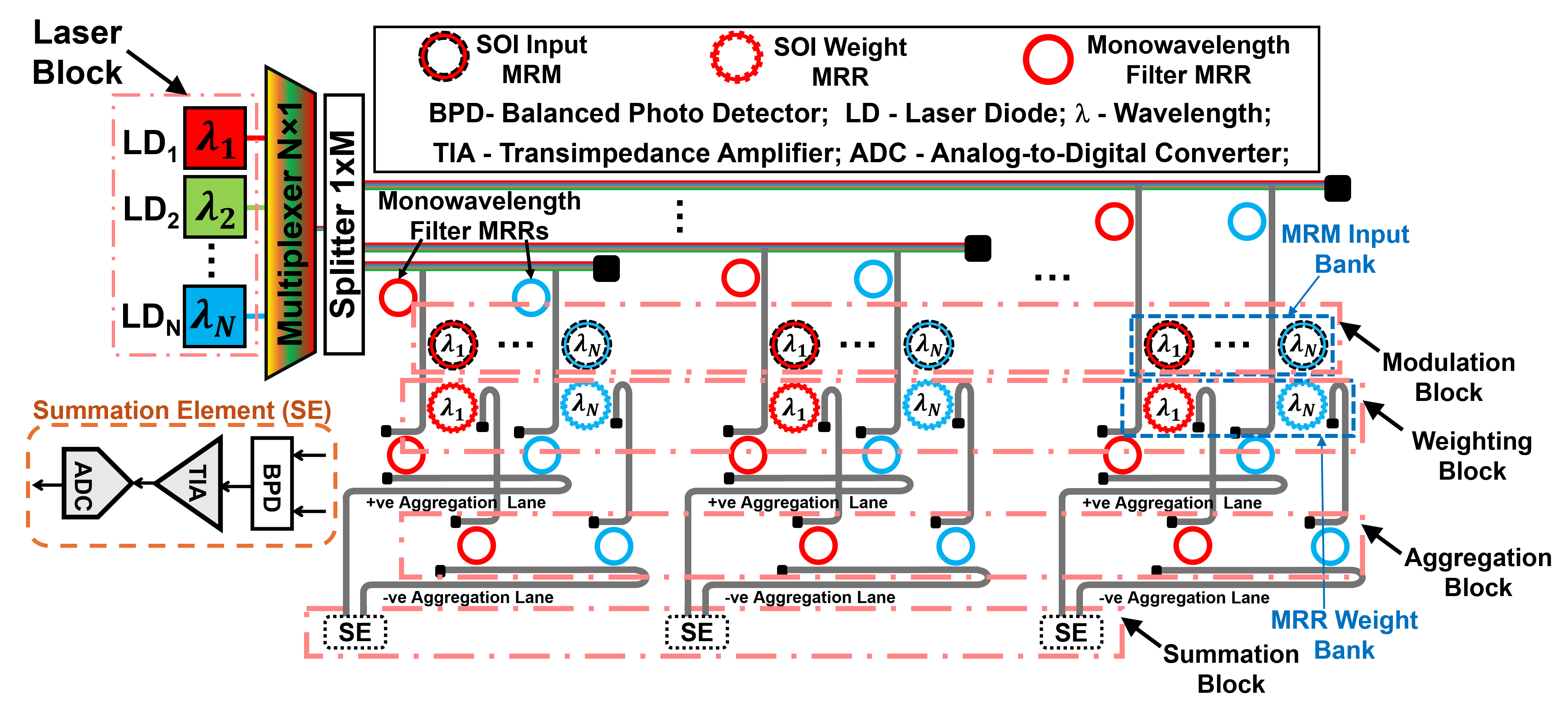}
    \caption{Illustration of the MWA organization of an SOI-based TPC.}
    \label{SOI_MWA}
\end{figure}

For detailed elucidation, Fig. \ref{SOI_MWA} illustrates the organization of an MWA TPC. As illustrated, the modulation and weighting blocks are placed before the aggregation block. In particular, the arrangement of each input-weight MRM pair is spectrally hitless \cite{karen2020proceeding} ensuring that each input-weight MRM pair produces a multiplication result and modulates this result onto a single-wavelength optical signal. This design eliminates inter-wavelength interference known as inter-modulation crosstalk \cite{padmaraju2014intermodulation}, at the MRMs, providing a notable advantage over the AMW and MAW organizations \cite{cases2022}. There are a total of \textit{M} DPEs in the TPC. And, in each DPE, there are a total of \textit{N} input-weight MRM pairs, with each MRM pair acting as a multiplier. The modulation and weighting blocks are connected to the aggregation block via a set of mono-wavelength filter MRRs. The aggregation block consists of positive and negative aggregation lanes that guide the signals to the SE block.

\subsection{Shortcomings of SOI Photonic GEMM Accelerators}\label{sec:shortcomings}
The SOI-based photonic GEMM accelerators face two major shortcomings that hinder their scalability, throughput, and energy efficiency.

\subsubsection{High Scattering Losses Due to High Index Contrast}
The high refractive index contrast between the silicon core (Si, 3.5) and the cladding (SiO$_{2}$, 1.5) in SOI-based waveguides serves as a double-edged sword. While it allows for the design of compact photonic waveguides by tightly confining the guided optical modes to the core, it also makes these waveguides highly susceptible to significant scattering losses. 
%mainly due to the following reasons. First, it increases the sidewall roughness of the waveguides arising from fabrication imperfections. Second,
This is because it leads to an enhanced interaction of the confined optical modes with the rough sidewalls of the waveguide. This enhanced mode-roughness interaction increases the scattering losses in the SOI waveguides. Studies have shown that even a slight RMS roughness of a few nanometers on the sidewalls, which is unavoidable due to fabrication imperfections, can result in substantial waveguide losses, often exceeding 3 dB/cm in SOI channel waveguides \cite{karempudi2023analysis,baets2016silicon}. 

\subsubsection{High Absorption Losses Due to Two-Photon Absorption}
The presence of free carriers in Si induces Two-Photon Absorption (TPA) in SOI-based photonic devices at telecom wavelengths. TPA increases the free-carrier density in the core (Si) material, leading to the Free-Carrier Absorption (FCA) effect in SOI waveguides \cite{baets2016silicon,karempudi2023analysis}. FCA results in higher absorption losses in SOI waveguides, particularly at elevated optical power levels. In DWDM applications, where multiple wavelengths are coupled into each SOI waveguide, the total optical power within the waveguide rises, triggering TPA-induced FCA effects and subsequently causing high absorption losses. Previous studies have shown that if the total number of wavelengths multiplexed into an SOI waveguide exceeds 20, the optical losses experienced by each additional wavelength channel propagating inside the waveguide increase by 0.1 dB/cm/wavelength \cite{lee2008ultrahigh,ophir2010demonstration}. 
%However, this increased demand for optical power diminishes a significant portion of the optical power budget, constraining the scalability and throughput of photonic GEMM accelerators. Furthermore, the need for higher input optical power diminishes the energy efficiency of these accelerators.

\subsection{Motivation}
To compensate for the high scattering and absorption losses in SOI waveguides, one option is to increase the input optical power. However, this increased input power whittles down a large portion of the optical power budget, leaving a small portion of the power budget available to support the scalability of size and spatial parallelism in SOI-based photonic GEMM accelerators. Additionally, the need for higher input optical power undermines the energy efficiency benefits associated with photonic GEMM accelerators. Therefore, there is a need for an alternative that can alleviate the optical signal losses and their detrimental impacts in photonic GEMM accelerators.

\section{SiNPhAR Architecture}
To alleviate the high scattering and absorption losses and related issues present in photonic GEMM accelerators, we propose to redesign photonic GEMM accelerators with the silicon nitride (SiN)-on-SiO$_2$ material system. Our idea is to address the root causes of high scattering and absorption losses in SOI-based designs, namely the high index contrast and TPA effect. The proposed SiN-on-SiO$_2$ material platform has been shown to exhibit ultra-low waveguide propagation losses (absorption + scattering losses) ($<$0.5 dB/cm \cite{ophir2010demonstration,baets2016silicon}) due to its low refractive index contrast compared to the SOI platform. In addition, the absence of free carriers in the SiN material eliminates the possibility of TPA-induced increase in absorption losses \cite{borghi2021modeling,baets2016silicon}. Our forged GEMM accelerator architecture based on the SiN-on-SiO$_2$ platform, which we call SiNPhAR architecture, is described in the following subsections. 

\subsection{Overview of SiNPhAR Tensor Processing Core (TPC)}
The main processing unit of our SiNPhAR architecture is a tensor processing core (TPC) (illustrated in Fig. \ref{SiNPhAR}), which follows the MWA TPC organization described in Section II.A, with several critical modifications in the constituent blocks. Across the modulation, aggregation, and weighting blocks, all the utilized photonic devices, including the waveguides, MRMs, and filter MRRs, are based on the SiN-on-SiO$_2$ material platform. We take the designs of the SiN-on-SiO$_2$ waveguides from \cite{blumenthal2018silicon} and filter MRRs from \cite{ilie2022thermo}, whereas we invent a new Indium Tin Oxide (ITO)-based all-pass design for SiN-on-SiO$_2$ MRMs (discussed in Section \ref{ITO-SiN-MRM}). In addition, as the summation (SE) block, we utilize our newly invented balanced photo-charge accumulator (BPCA) (discussed in Section \ref{bpca}). Atop these modifications, a SiNPhAR TPC employs all-pass MRMs in its weighting blocks, which is different from the add-drop MRRs used in the weighting blocks of SOI-based MWA TPC. Because of this difference, a SiNPhAR TPC utilizes a filter MRR after each input-weight MRM pair. This filter MRR allows routing of the optical signal incoming from the input-weight MRM pair onto the positive or negative aggregation lanes, depending on the sign of the multiplication result produced by the input-weight MRM pair. The structure and functionality of various blocks of a SiNPhAR TPC are discussed in the upcoming subsections.

\begin{figure}[h!]
    \centering
    \includegraphics[width = \linewidth]{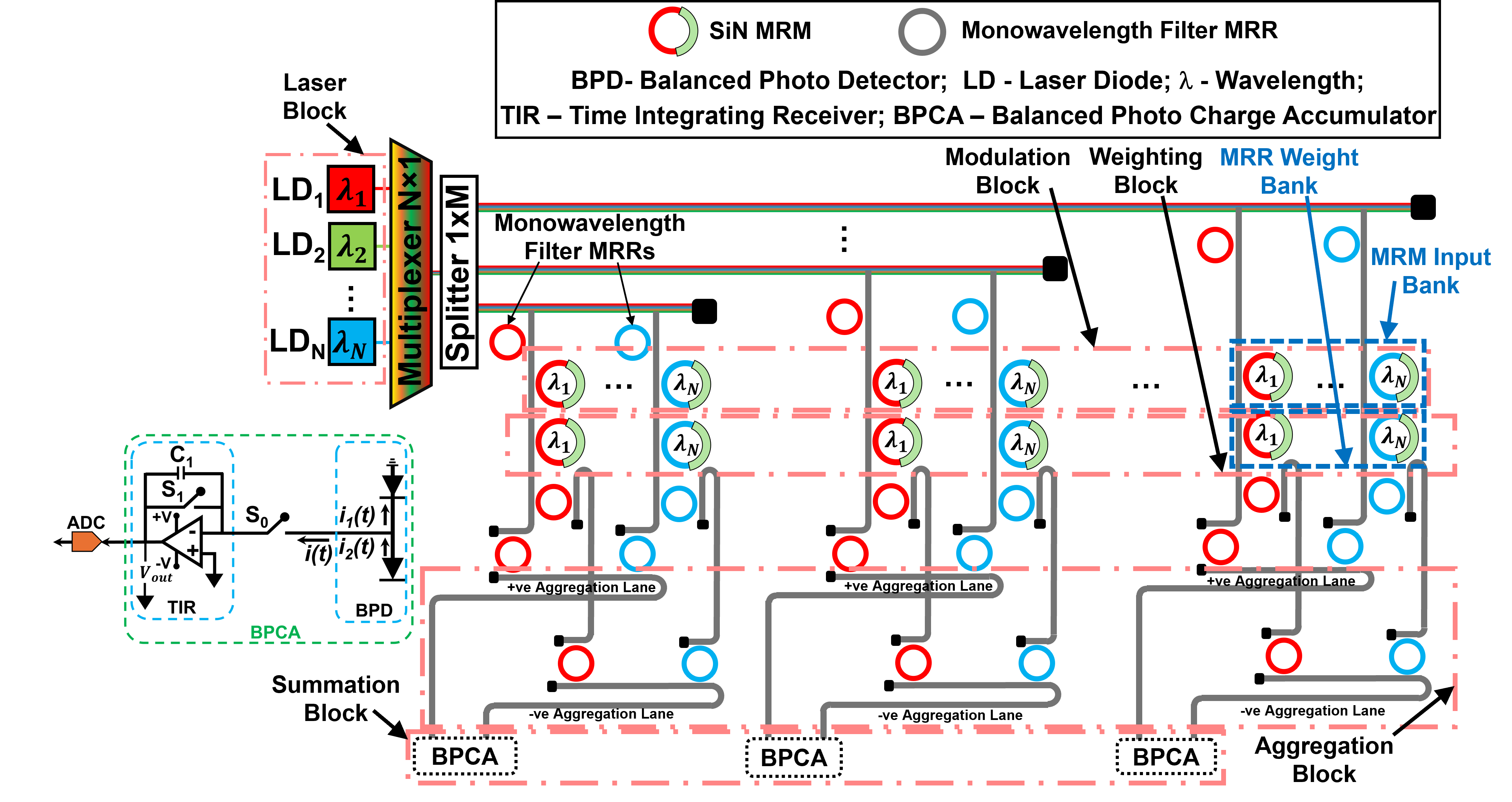}
    \caption{Schematic of a TPC of our SiNPhAR GEMM Accelerator.}
    \label{SiNPhAR}
\end{figure}

\subsection{ITO-Based SiN MRM for Input Encoding}\label{ITO-SiN-MRM}
\textbf{Structure.} Fig. \ref{Fig:1}(a) and Fig. \ref{Fig:1}(b), respectively, show the top-view and cross-sectional schematic of our SiN-on-SiO$_2$ MRM. Fig. \ref{Fig:1fab} shows the scanning electron microscope (SEM) image of our fabricated SiN-on-SiO$_2$ MRM. The active region in the upper cladding of the MRM consists of a stack of two ITO thin films with a silicon dioxide (SiO$_2$) thin film in between (i.e., an ITO-SiO$_2$-ITO stack). From Fig. \ref{Fig:1}(b), we have a 300 nm thick SiN-based MRM waveguide, two 10 nm thick ITO films, and 15 nm thick SiO$_2$ layer. Upon applying voltage across the ITO-SiO$_2$-ITO stack (through the Au pads shown in Fig. \ref{Fig:1}(a) and Fig. \ref{Fig:1fab}), free carriers accumulate in the ITO films at the ITO-SiO$_2$ interfaces for up to ~5 nm depth in the ITO films \cite{chrostowski2014}, making these accumulation regions in the ITO films high-carrier-density active regions. In these regions, a free-carriers-assisted, large-amplitude modulation in the permittivity and refractive index of the ITO material has been previously reported \cite{chrostowski2014}. This kind of free-carriers based index modulation in the ITO films of our MRM follows the Drude-Lorentz model from \cite{ma2015indium}. Accordingly, as the carrier concentration in the ITO accumulation regions increases, the refractive index of the ITO films decreases. As a result, the effective refractive index of our SiN-on-SiO$_2$ MRM from Fig. \ref{Fig:1fab} also decreases, causing a blue shift in its resonance wavelength that in turn causes a transmission modulation at the through port of the MRM. The electro-refractive activity of our SiN-on-SiO$_2$ MRM is confined only in the ITO-SiO$_2$-ITO cladding. This is different from the Si-SiO$_2$-ITO capacitor-based MRR modulator from \cite{li2019silicon}, which has the electro-refractive as well as electro-absorptive activities in both its Si-based MRR core and SiO$_2$-ITO based cladding.

\begin{figure}[h!]
    \centering
    \includegraphics[width = \linewidth]{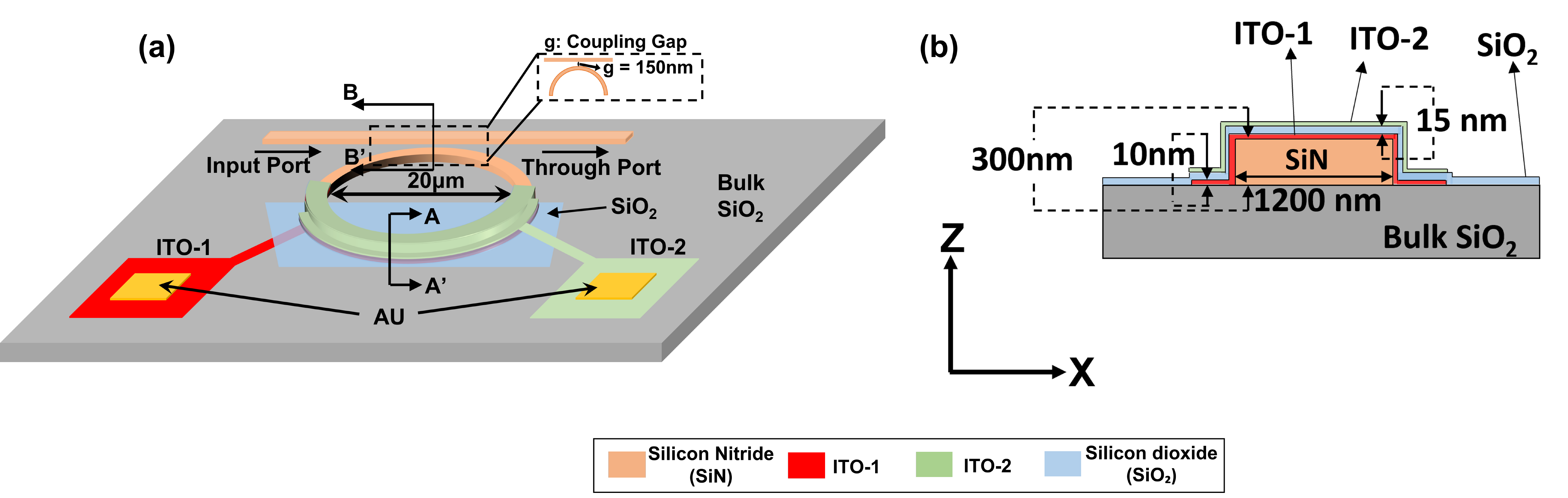}
    \caption{(a) Top view, (b) Cross-sectional view (along AA') of our SiN-on-SiO$_2$ MRM.}
    \label{Fig:1}
\end{figure}

\begin{figure}[h!]
    \centering
    \includegraphics[width = 165pt]{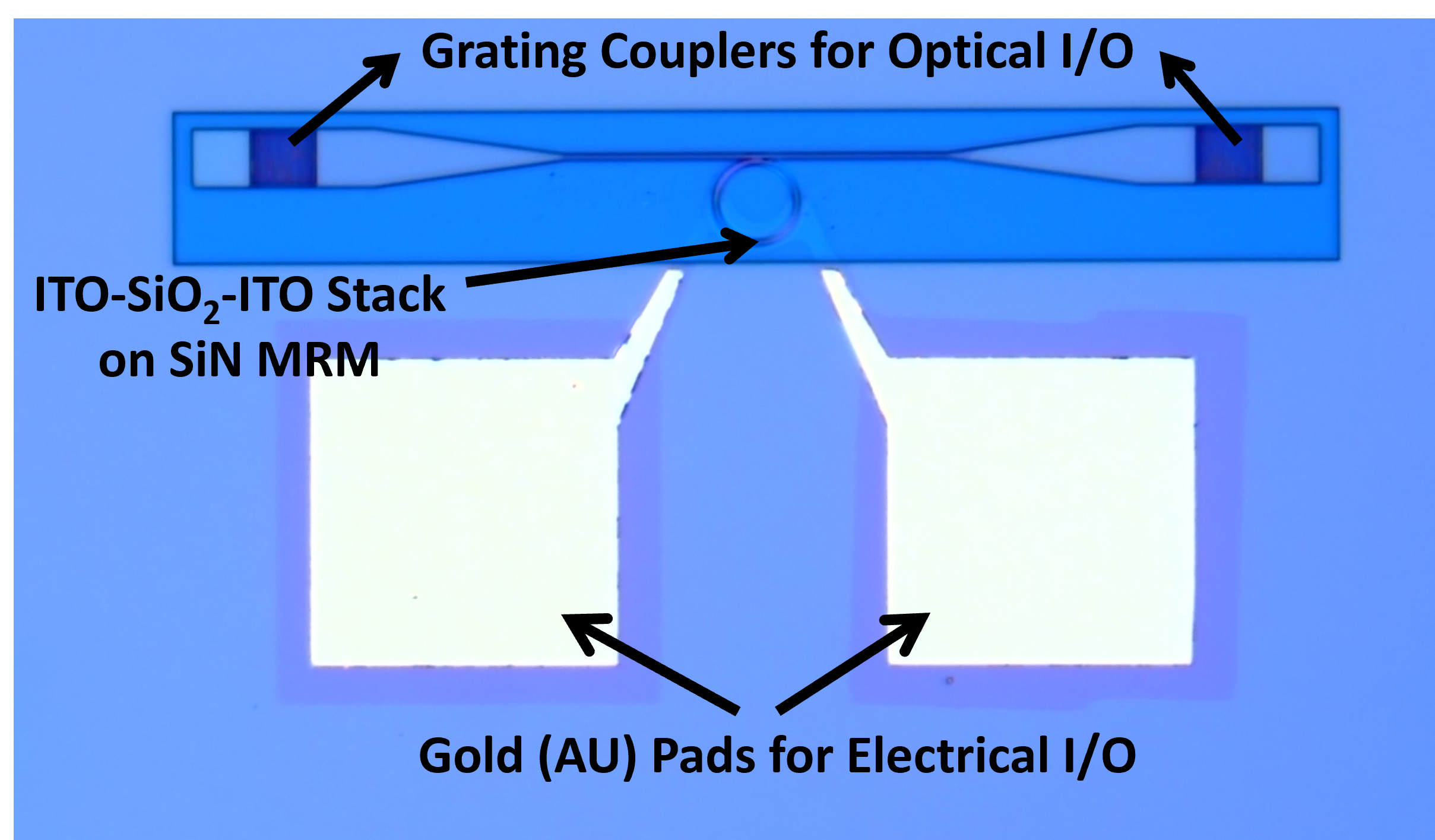}
    \caption{SEM image of our fabricated SiN-on-SiO$_2$ MRM.} 
    \label{Fig:1fab}
\end{figure}

\begin{table}[h!]
\caption{Free-carrier concentration (N), real index (Re($\eta_{ITO}$)), and imaginary index (Im($\eta_{ITO}$)) for the ITO accumulation layer in our MRM. The real and imaginary effective index (Re($\eta_{eff}$), Im($\eta_{eff}$)), operating voltage (V), and induced resonance shift ($\Delta\lambda_{r}$) for our MRM.}
\begin{tabular}{|c|c|c|c|c|c|c|}
\hline
\begin{tabular}[c]{@{}c@{}}N \\ ($cm^{-3}$)\end{tabular} & \begin{tabular}[c]{@{}c@{}}Re\\ ($\eta_{ITO}$)\end{tabular} & \begin{tabular}[c]{@{}c@{}}Im\\ ($\eta_{ITO}$)\end{tabular} & \begin{tabular}[c]{@{}c@{}}Re\\ ($\eta_{eff}$)\end{tabular} & \begin{tabular}[c]{@{}c@{}}Im\\ ($\eta_{eff}$)\end{tabular} & V   & \begin{tabular}[c]{@{}c@{}} $\Delta\lambda_{r}$\\ (pm)\end{tabular} \\ \hline
1 × $10^{19}$                                            & 1.9556                                              & 0.0100                                              & 1.9735                                              & 0.0001                                              & 0   & 0                                                  \\ \hline
5 × $10^{19}$                                            & 1.9111                                              & 0.0403                                              & 1.9724                                              & 0.0003                                              & 1.8 & 830                                                \\ \hline
9 × $10^{19}$                                             & 1.8667                                              & 0.0896                                              & 1.9712                                              & 0.0006                                              & 3.7 & 1580                                               \\ \hline
13 × $10^{19}$                                           & 1.8222                                              & 0.1289                                              & 1.9701                                              & 0.0011                                              & 5.5 & 2470                                               \\ \hline
17 × $10^{19}$                                            & 1.7778                                              & 0.1582                                              & 1.9692                                              & 0.0017                                              & 7.3 & 3210                                               \\ \hline
20 × $10^{19}$                                            & 1.7333                                              & 0.1874                                              & 1.9680                                              & 0.0022                                              & 9.2 & 4000                                               \\ \hline
\end{tabular}
\label{Table:1}
\end{table}

\textbf{Testing and Characterization Results.} To test and characterize our MRM, we evaluated the required voltage levels across the Au pads (Figs. \ref{Fig:1}(a) and \ref{Fig:1fab}) for achieving various free-carrier concentrations in the ITO films. Then, we extracted the corresponding ITO index change values for various free-carrier concentrations. These results are listed in Table \ref{Table:1}. We also extracted the effective index change and transmission spectra of our MRM (shown in Table \ref{Table:1} and Fig. \ref{Fig:2} respectively) at various applied voltages for the operation around 1.6 $\mu$m wavelength (L-band). From Fig. \ref{Fig:2}, our MRM achieves a resonance shift of up to 4 nm upon applying 9.2 V across the thin-film stack, which renders the resonance tuning (modulation) efficiency of $\sim$450 pm/V. This is a crucial outcome, as our MRM has a relatively very low overlap between the optical mode and free-carrier perturbation (only ~10\% of the guided optical mode overlaps with the upper cladding based on ITO) compared to the silicon ITO-based modulators (e.g., \cite{li2019silicon}).

Further, from the spectra in Fig. \ref{Fig:2}, we evaluate the FSR of our MRM to be $\sim$18 nm. We evaluated the insertion loss and loaded Q-factor of our MRM to be $\sim$0.235 dB and $\sim$2000 respectively. We also evaluated the capacitance density of the ITO thin-film stack covering the MRM rim to be $\sim$2.3 fF$/\mu$m$^{2}$ for the 15 nm thick SiO$_2$ layer. Moreover, we measured the optical eye diagrams for the MRM at 30 Gb/s and 55 Gb/s operating bitrates and found that our MRM can achieve 8.2 dB extinction ratio for OOK modulation at 30 Gb/s bitrate, which confirms the high-speed operation of our MRM.

\begin{figure}[h!]
    \centering
    \includegraphics[width = 260pt]{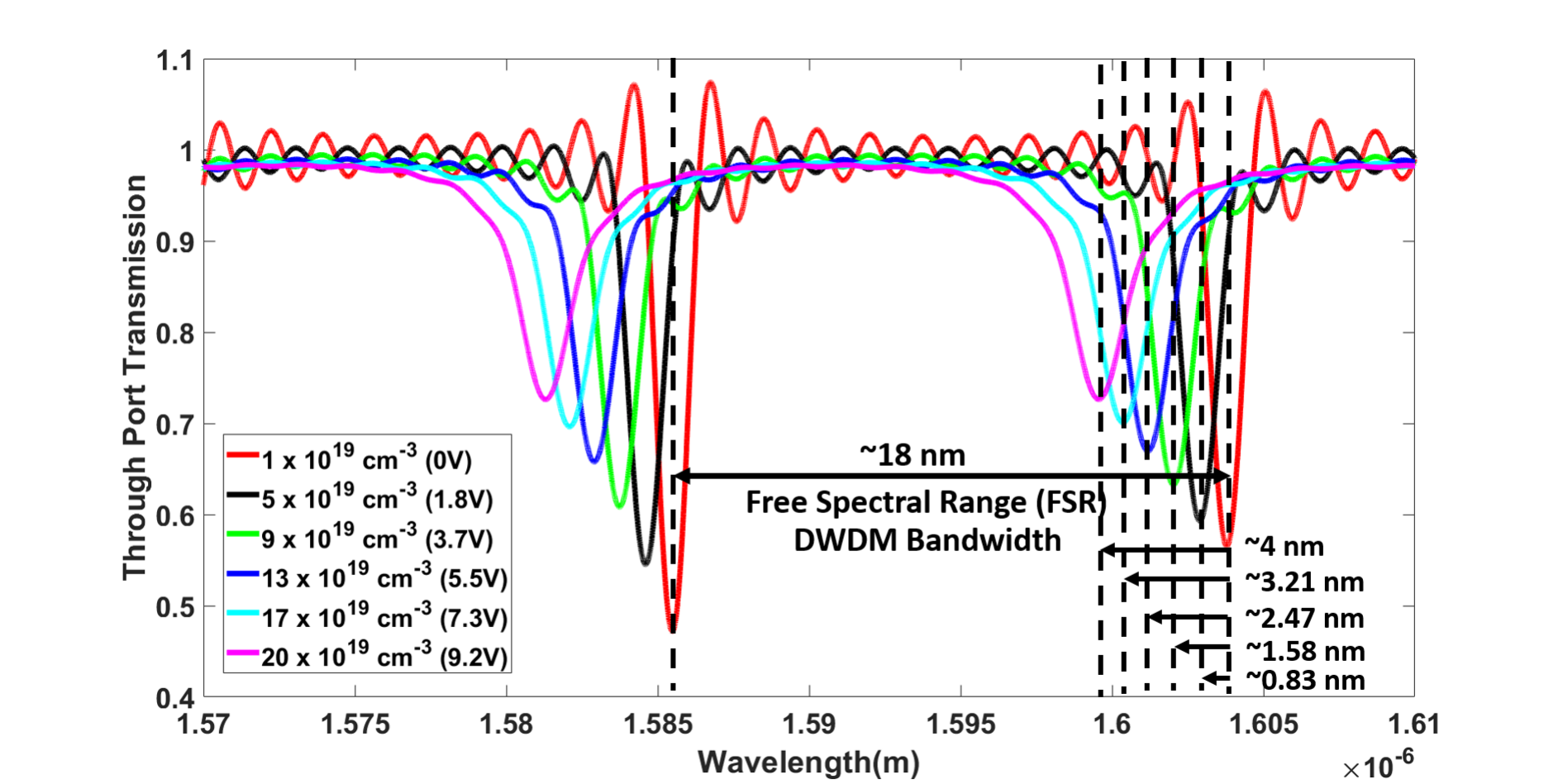}
    \caption{Transmission spectra of our SiN-on-SiO$_2$ MRM.}
    \label{Fig:2}
\end{figure}

\textbf{Use as an Input-Encoding Element.} As described above, our MRM can produce a high-speed optical signal. This signal can be generated as a temporal train of optical amplitude symbols, similar to how an SOI MRM is used in SOI-based photonic GEMM accelerators to generate an optical signal as a temporal train of optical symbols \cite{deapcnn}. In this signal, the amplitude of each symbol represents an analog input value. Thus, our MRM, when used for input encoding, can produce an optical signal as a temporal train of analog input values.  

\subsection{ITO-Based SiN MRM for Weighting}\label{SiN-Weighting}
Our unique ITO-based SiN-on-SiO$_{2}$ MRM serves a dual purpose in our TPC, functioning not only as a high-speed electro-optic input encoding element but also as a high-speed electro-optic weighting element. In this paper, we explored its application in performing precise weighting of input-modulated optical signals. To assess the effectiveness of our MRM in this context, we conducted a comprehensive study with weighting values of 3-bit resolution.

Intuitively, a weighting MRM of 3-bit (4-bit) resolution should be able to alter the transmission of an input optical amplitude/symbol to one of the 2$^{3}$=8 (2$^{4}$=16) distinct output amplitude levels. These 8 or 16 distinct output amplitude levels are achieved in our MRM at its through port by enabling electro-optic shifting of its resonance passband to 8 or 16 distinct spectral locations. Consequently, to imprint a certain 3-bit or 4-bit weighting on an input optical symbol, the input optical symbol is applied at the input port of the MRM, and then, the analog-converted (via a digital-to-analog converter) 3-bit or 4-bit weight value is applied to the electrical I/O pads of the MRM (see Fig. \ref{Fig:1fab}) to effect an electro-optic shifting of the MRM's resonance passband. The shifted passband programs the through-port transmission of the input optical symbol to a corresponding output amplitude value from the 8 or 16 possible output amplitude levels. Fig. \ref{Weighting} illustrates how the shifting of our MRM's resonance passband enables weighting with 3-bit resolution. In the figure, $\lambda_T$ shows the optical wavelength carrying the input optical symbol, and the 8 resonance passbands show the electro-optically shifted spectral locations corresponding to 8 output transmission amplitudes. When the spectral position of the passband is shifted, the intersection point of the passband with $\lambda_T$ changes, which in turn alters the transmission amplitude. Thus, our MRM can be used to implement a weighting of an input optical symbol. 

This operation of MRM weighting element can also be used to weight a high-speed optical signal output from an input-encoding MRM. For that, the MRM resonance passband is shifted to achieve different weighting amplitudes at a speed that is matched to the symbol rate of the input high-speed optical symbol. As a result, each symbol of the input optical signal is weighted with a unique weighted value to generate a weighted optical signal. Each symbol of the weighted optical signal, thus, represents a multiplication result (product) between the input and weight values. Therefore, the weighted optical signal generated by a weighting MRM is also referred to as an optical product signal. Each symbol of this optical product signal, depending on its sign, is routed to the positive or negative aggregation lane, in the aggregation block of the SiNPhAR TPC.

%of the    e3-bit 0r 4-bit For that, having the using From the simulations, we obtained transmission spectra for 8 distinct amplitude levels, representing a total of 2$^{3}$ possible magnitudes for a bit resolution of 3-bits. These spectra were extracted at the through port of the MRM, as depicted in Figure \ref{Weighting}. To ensure accuracy, we meticulously selected unique target transmission values corresponding to each amplitude level. These values were determined at the points where the spectra intersected with the carrier wavelength $\lambda_T$, as shown in Figure \ref{Weighting}. This careful selection process guaranteed that each amplitude level (or weight value) corresponded to a specific and unique target transmission value associated with a particular transmission spectrum and carrier wavelength intersection. Utilizing these unique transmission values obtained at different amplitudes, we successfully implemented the weighting of the input-encoded wavelength channel. Through this process, the intensity of the wavelength channel accurately represents the multiplication of the corresponding input and weight operands. 

\begin{figure}[h!]
    \centering
    \includegraphics[width = 230pt]{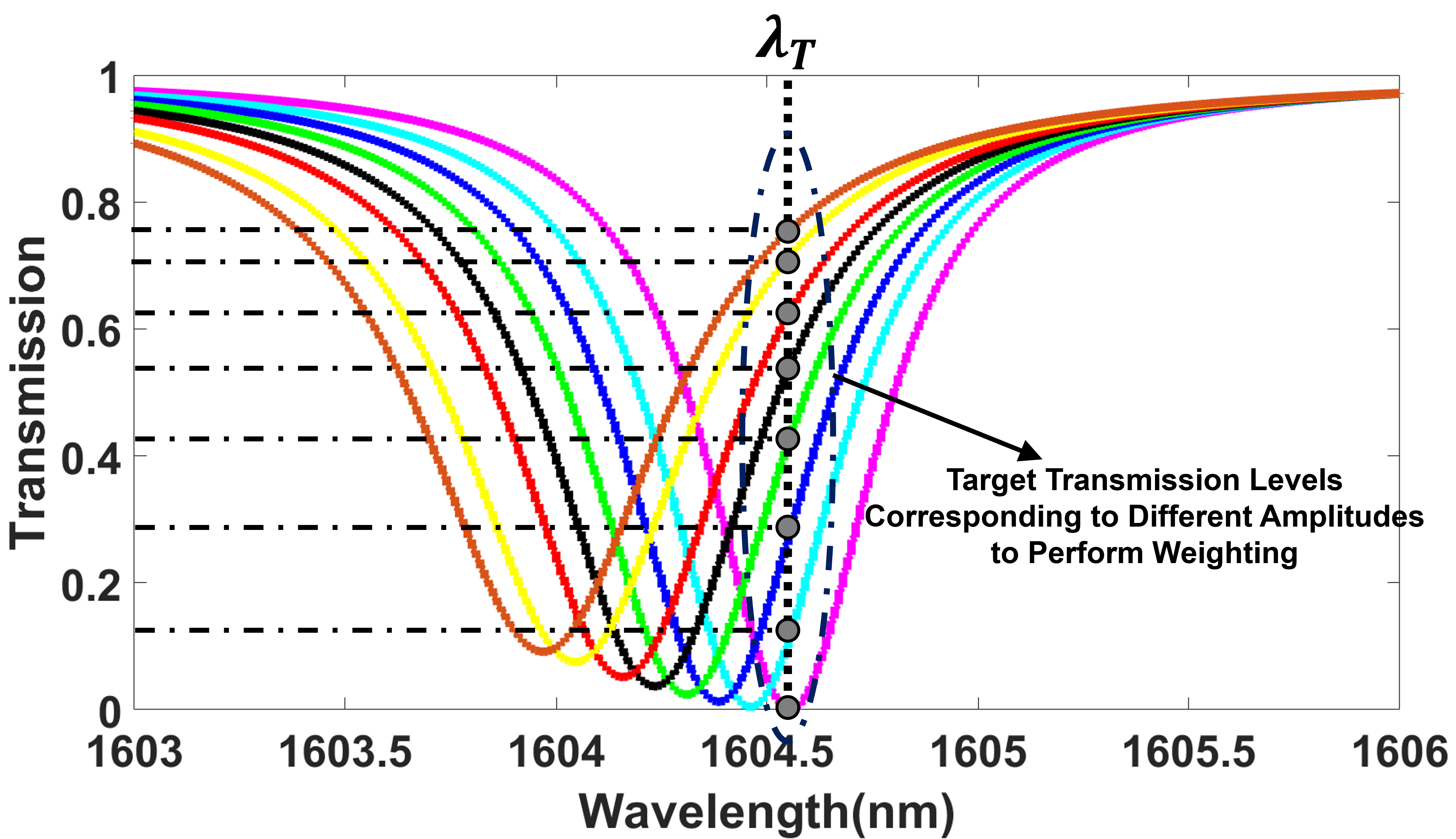}
    \caption{Transmission spectra measured at the through port of our SiN-on-SiO$_2$ MRM weighting element for various transmission amplitudes. These different transmission amplitudes at $\lambda_{T}$ signify the weighting of the input optical amplitude symbol.}
    \label{Weighting}
\end{figure}

\subsection{Summation with Balanced Photo-Charge Accumulator (BPCA)}\label{bpca}
In a DPE of a SiNPhAR TPC, each weighting MRM outputs one optical product signal, and a total of \textit{N} such optical product signals are aggregated into the positive and negative aggregation lanes. The aggregation lanes deliver these optical product signals to the BPCA circuit for summation. Our BPCA circuit is collectively inspired by the time integrating receiver (TIR) design from \cite{alexandermit2022} and the photodetector-based optical pulse/symbol accumulator design from \cite{BhaskaranPCA2022}. As illustrated in Fig. \ref{SiNPhAR}, a BPCA circuit employs two photodiodes, each connected to the positive and negative aggregation lanes. These photodiodes are interlinked in a balanced configuration, commonly referred to as a balanced photodiode (BPD) configuration. The BPD is connected to a TIR. The TIR comprises an amplifier and a feedback capacitor/switch pair (Fig. \ref{SiNPhAR}). 

%It functions as a current-to-voltage converter circuit by integrating the incoming electrical current over a period.

A total of \textit{N} optical product symbols arrive at the BPCA during a symbol cycle. The constituent BPD of the BPCA performs an incoherent superposition (signed summation) of all these \textit{N} optical product symbols received within that cycle. Consequently, the incoherent superposition first enables the creation of a net optical symbol. The total optical energy packetized within a new optical symbol is proportional to the signed summation of the \textit{N} optical product symbols. The BPD transduces this net optical symbol into a balanced photocurrent symbol, which is further transduced by the TIR of the BPCA into an analog voltage level accrued on the capacitor of the TIR. This accused analog voltage level, thus, represents a summation of \textit{N} products, i.e., an \textit{N}-sized dot product. 

This \textit{N}-sized dot product result, in the form of the accrued analog voltage level, can be held by the TIR. As new net optical symbols keep arriving in subsequent symbol cycles, the TIR enables a gradual integration (temporal accumulation) of the individual dot product results over multiple symbol cycles to generate a larger ($>$\textit{N}-sized) dot product result. This is possible because the \textit{N}-sized dot product results arriving at the TIR can sequentially charge the TIR's capacitor so that the net accumulated charge and, consequently, the net analog voltage accrued on the capacitor over multiple symbol cycles provides the signed sum of the individual dot product results. This final sum value in the analog voltage format can be sampled and sent to the analog-to-digital converter (ADC) for conversion in the binary format. Thus, the BPCA of a SiNPhAR DPE can essentially enable the processing of very large-sized ($>$\textit{N}-sized) dot products.

\section{Evaluation and Discussion}
\subsection{Scalability Analysis}
To perform scalability analysis, we utilized the equations provided in \cite{al2022scaling}, reproduced as Eqs. \ref{eq1}, \ref{eq2} and \ref{eq3}. The parameters and their corresponding values \cite{al2022scaling,cases2022,vatsavai2023sconna} required to solve these equations are listed in Table \ref{Table:2}. We devised a two-step procedure to determine the optimal value of \textit{N} and \textit{M} (\textit{N} refers to the count of input-weight MRM pairs per DPE, whereas \textit{M} refers to the count of DPEs per TPC) for a given bit precision and data rate (DR), as outlined below.

\textbf{Step 1.} We calculate the photodiode (PD) sensitivity by solving Eq. \ref{eq1} for the specified bit precision and DR.

\textbf{Step 2.} Next, we perform an exhaustive search to find the optimal value of \textit{N} (assuming \textit{N} = \textit{M}) for the specified bit precision and DR, using Eqs. \ref{eq2} and \ref{eq3}. In this step, we solve Eq. \ref{eq3}, which represents the error function (\textit{ef}). The \textit{ef} is the difference between the optical power reaching the photodiode (P$_{output}$), calculated from Eq. \ref{eq2}, and the PD sensitivity obtained in Step 1/Eq. \ref{eq1}. We sweep for different values of \textit{N}, and the optimal value of \textit{N} for the specified bit precision, and DR is the one for which the \textit{ef} yields the minimum positive value. Notably, when solving Eqs. \ref{eq2} and \ref{eq3}, we consider P$_{inc}$ (see Table \ref{Table:2} for definition) to be zero for '\textit{N}' values less than 20 wavelengths/waveguide. However, beyond 20 wavelengths/waveguide, we account for a changing P$_{inc}$ between the SOI and the SiN waveguides, as reported in Table \ref{Table:2}. This is because the TPA-induced absorption losses in the SOI waveguide substantially increase if the total number of multiplexed wavelengths in an SOI waveguide exceeds 20 (as discussed in Section \ref{sec:shortcomings}). In contrast, this phenomenon is not observed in SiN waveguides, as detailed in Section III.

\begin{figure}[H]
    \centering
    \includegraphics[width=\linewidth]{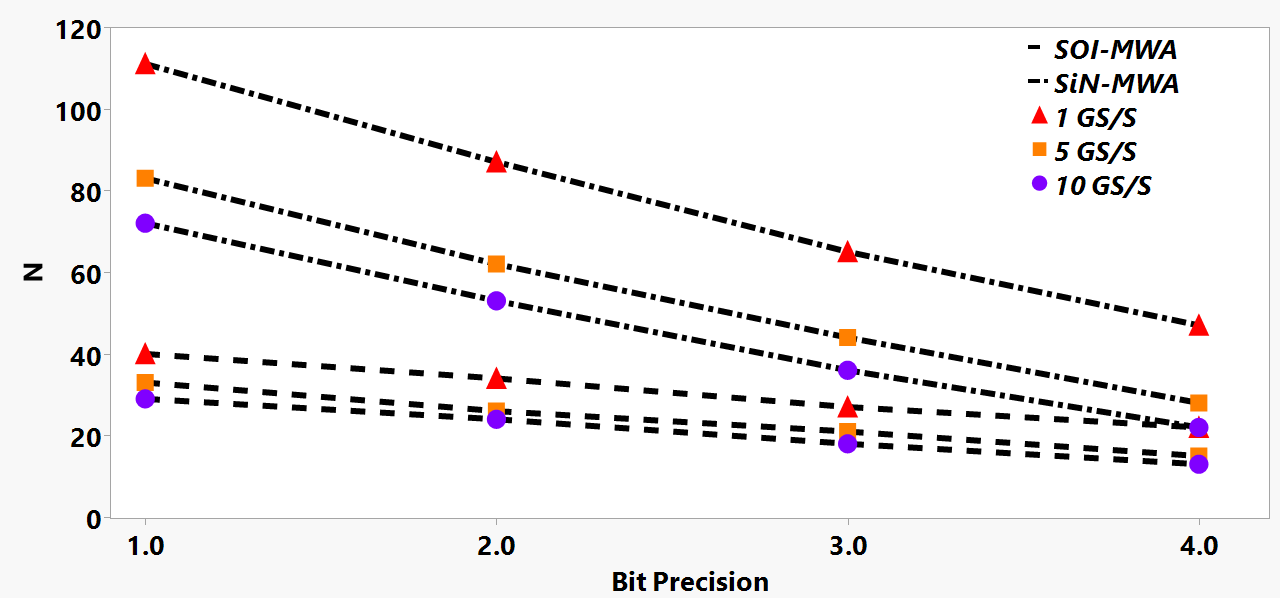}
    \caption{Supported TPC size N(=M) for bit precision = \{1,2,3,4\}-bits at data rates (DRs)=\{1,5,10\} GS/s, for SOI-MWA TPC and SiNPhAR TPC.}
    \label{scalability}
\end{figure}
For our analysis, we considered bit-precision values ranging from 1-bit to 4-bits and a set of DRs namely 1GS/S, 5GS/S, and 10GS/S. The results of our analysis are illustrated in Fig. \ref{scalability}. In addition to our SiNPhAR, we conducted the scalability analysis for an SOI-based MWA TPC (Fig. \ref{SOI_MWA}) named as \textit{SOI-MWA}. From Fig. \ref{scalability}, our SiNPhAR can support a larger value of \textit{N} compared to SOI-MWA. For instance, our SiNPhAR can support \textit{N=52} for a bit-precision of 3-bits, DR = 1GS/S, and an input laser power of 10dBm, which is larger compared to SOI-MWA that supports \textit{N=35}. This advantage primarily stems from the reduced propagation losses in the SiN-on-SiO${_2}$ waveguides and the lower insertion loss of the ITO-based SiN-on-SiO${_2}$ MRMs in SiNPhAR, compared to the SOI waveguides and MRMs in SOI-MWA. Consequently, this creates a larger room in the optical power budget, allowing for the accommodation of a larger \textit{N} in our SiNPhAR TPC.

\begin{figure*}
\centering
\begin{equation*}
B = \frac{1}{6.02} \left[20 \log_{10} \left( \frac{RP_{\text{PD-opt}}}{\left( \sqrt{2q \left( RP_{\text{PD-opt}} + I_d \right) + \frac{4KT}{R_L} + \left( RP_{\text{PD-opt}} \right)^2 \text{RIN}} + \sqrt{2qI_d + \frac{4KT}{R_L}} \right) \sqrt{\frac{DR}{\sqrt2}}}\right) - 1.76 \right]
\tag{1}
\label{eq1}
\end{equation*}
\end{figure*}

\begin{figure*}
\centering
\begin{equation}
\begin{aligned}
P_{\text{output}}(\text{dBm}) = & P_{\text{L}} - P_{\text{SMF}} - P_{\text{C}} - (P_{\text{WG-IL}} \times d_{\text{MRR}} \times N) - (P_{\text{Inc}} \times d_{\text{MRR}} \times (\text{N-20})) - (P_{\text{sp}} \times \log_{2}(\text{N})) - P_{\text{MRM}} - P_{\text{MRR}} \\ 
& - ((\text{N-1}) \times P_{\text{MRM-OBL}}) - ((\text{N-1}) \times P_{\text{MRR-OBL}}) - P_{\text{penalty}}
\end{aligned}
\tag{2}
\label{eq2}
\end{equation}
\end{figure*}

\begin{figure}
\centering
\begin{equation*}
\text { ef }(\mathrm{B}, \mathrm{DR}, \mathrm{N})=P_{\text {output }}(\mathrm{N})-P_{P D-\text { opt }}(\mathrm{B}, \mathrm{DR})
\tag{3}
\label{eq3}
\end{equation*}
\end{figure}

\begin{table}[h!]
\centering
\caption{Definition and values of various parameters used in Eq. \ref{eq1}, Eq. \ref{eq2}, and Eq. \ref{eq3} (from \cite{al2022scaling}) for the scalability analysis.}
\begin{tabular}{|c|c|c|}
\hline
Parameter                 & Definition                                                                                                                   & Value                                     \\ \hline
P$_{L}$                        & Laser Power Intensity                                                                                                        & 10 dBm                                    \\ \hline
P$_{SMF}$                      & \begin{tabular}[c]{@{}c@{}}Attenuation by the \\ Single Mode Fiber\end{tabular}                                              & 0 dB                                      \\ \hline
P$_{C}$                        & \begin{tabular}[c]{@{}c@{}}Fiber-to-Chip \\ Coupling Insertion Loss\end{tabular}                                             & 1.6 dB                                    \\ \hline
\multirow{2}{*}{P$_{WG-IL}$}   & \begin{tabular}[c]{@{}c@{}}Propagation Loss of \\ SOI Waveguide\end{tabular}                                              & 1.5 dB/cm        \\ \cline{2-3} 
                          & \begin{tabular}[c]{@{}c@{}}Propagation loss of \\ SiN Waveguide\end{tabular}                                                 & 0.5 dB/cm        \\ \hline
\multirow{2}{*}{P$_{inc}$}     & \begin{tabular}[c]{@{}c@{}}Increase in Propagation Loss of the \\ SOI waveguide atop 20 $\lambda$s/waveguide\end{tabular}   & 0.1 dB/cm/$\lambda$                       \\ \cline{2-3} 
                          & \begin{tabular}[c]{@{}c@{}}Increase in Propagation Loss of the \\ SiN waveguide atop 20 $\lambda$s/waveguide\end{tabular} & 0.01 dB/cm/$\lambda$                     \\ \hline
P$_{SP}$                       & Splitter Insertion Loss                                                                                                      & 0.01 dB                                   \\ \hline
\multirow{2}{*}{P$_{MRM}$}     & \begin{tabular}[c]{@{}c@{}}Transmission Insertion Loss \\ of the SOI MRM\end{tabular}                                        & 4 dB                                      \\ \cline{2-3} 
                          & \begin{tabular}[c]{@{}c@{}}Transmission Insertion Loss    \\ of the SiN MRM\end{tabular}                                     & 0.235 dB                                  \\ \hline
P$_{MRR}$                      & \begin{tabular}[c]{@{}c@{}}Transmission Insertion Loss \\ of the SOI MRR\end{tabular}                                        & 0.01 dB                                   \\ \hline
P$_{MRM-OBL}$                  & \begin{tabular}[c]{@{}c@{}}Out-of-Band Insertion \\ Loss (OBL) of the MRM\end{tabular}                                       & 0.01 dB                                   \\ \hline
P$_{MRR-OBL}$                  & \begin{tabular}[c]{@{}c@{}}Out-of-Band Insertion \\ Loss (OBL) of the MRR\end{tabular}                                       & 0.01 dB                                   \\ \hline
\multirow{2}{*}{P$_{Penalty}$} & \begin{tabular}[c]{@{}c@{}}Network Penalty \\ for SOI-MAW\end{tabular}                                                       & \multirow{2}{*}{1.8 dB}                   \\ \cline{2-2}
                          & \begin{tabular}[c]{@{}c@{}}Network Penalty \\ for SiNPhAR\end{tabular}                                                       &                                           \\ \hline
R                         & PD Responsivity                                                                                                              & 1.2                                       \\ \hline
q                         & Charge of an Electron (C)                                                                                                        & 1.6 $\times$ 10$^{-19}$    \\ \hline
I$_{d}$                        & PD Dark Current                                                                                                              & 35 nA                                     \\ \hline
K                         & Boltzmann Constant (J/K)                                                                                                           & 1.38 $\times$ 10$^{-23}$ \\ \hline
T                         & Absolute Temperature (K)                                                                                                        & 300                                     \\ \hline
R$_{L}$                        & Load Resistance (Ohms)                                                                                                              & 50                                   \\ \hline
RIN                       & Relative Intensity Noise (dB/Hz)                                                                                                     & -140                                \\ \hline
B                         & Bit-Precision                                                                                                                &                                           \\ \hline
P$_{PD-OPT}$                   & PD Sensitivity                                                                                                               & -                                         \\ \hline
\end{tabular}
\label{Table:2}
\end{table}

\subsection{System-Level Evaluation Method}
\subsubsection{System-Level Implementation}
Fig. \ref{systemlevelimplement} illustrates the general system-level implementation of a photonic GEMM accelerator. It consists of global memory that stores convolutional neural network (CNN) parameters and a pre-processing and mapping unit. It has a mesh network of tiles. Each tile contains 4 dot-product units (DPUs) (a DPU is synonymous/analogous to a TPC) interconnected (via H-tree) with a unified buffer as well as pooling and activation units. Each TPC/DPU consists of multiple DPEs and each DPE is equipped with a dedicated input and output FIFO buffer \cite{Wang2021-ISCAS} to store intermittent weights, inputs, and partial sum values. The generic DPUs/TPCs in the system are replaced with SiNPhAR TPCs and SOI-MWA TPCs, respectively, to derive SiNPhAR and SOIPhAR accelerator system architectures.

\subsubsection{Simulation Setup}
In our study, we employed a custom Python-based simulator to emulate the system-level deployment of SiNPhAR and SOIPhAR accelerator architectures. The simulation involved the inference of four distinct CNN models (with a batch size of 1): ShuffleNet V2 \cite{shufflenet}, GoogleNet \cite{googlenet}, and ResNet50 \cite{resnet}. We converted the convolutional layers and fully connected layers of these CNNs into GEMM operations using the Toeplitz matrix transformations or im2col functions \cite{cases2022,vatsavai2023sconna}, and then accelerated these GEMM operations on our considered accelerators. We conducted a comparative analysis of SiNPhAR and SOIPhAR accelerator architectures in the context of inferring 8-bit integer quantized CNN models. Key metrics such as Frames per second (FPS) and FPS/W (energy efficiency) were evaluated. All accelerators were operated across data rates of 1GS/s, 5GS/s, and 10GS/s. Each TPC was operated at 4-bit precision; therefore, two TPCs were used with back-end shift-and-add circuits to achieve 8-bit computational precision. For these specific data rates, SOIPhAR and SiNPhAR achieve \textit{N} (TPC size) as shown in Table \ref{Table:3}. Our evaluation is based on output stationary dataflow. To ensure a fair comparison, we carried out an area proportionate analysis, wherein we adjusted the TPC count for each SiNPhAR and SOIPhAR variants listed in Table \ref{Table:3} so that the total area consumption of all TPCs per variant remained constant across all variants.

\begin{figure}[h]
  \centering
  \includegraphics[scale=0.35]{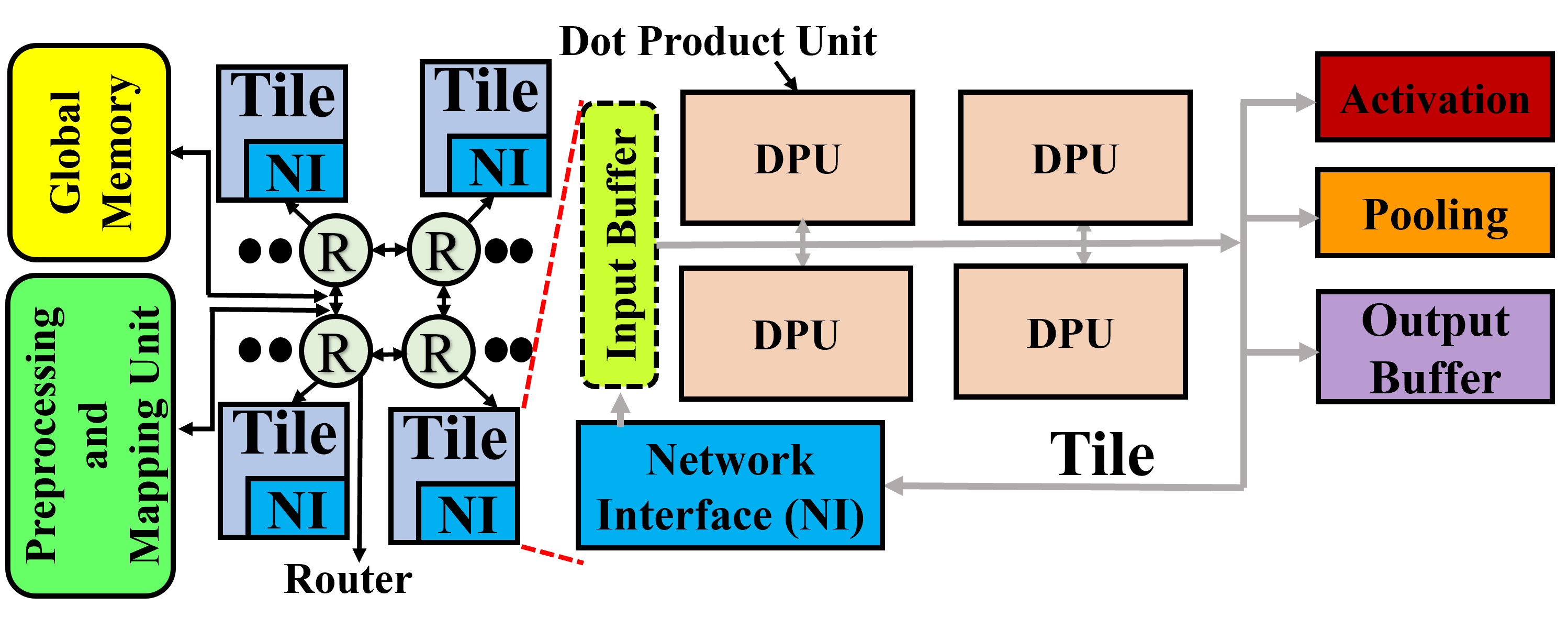}
  \caption{System-level implementation of SiNPhAR accelerator. DPU=TPC.} 
  \label{systemlevelimplement}

\end{figure}

Table \ref{Table:2} outlines the parameters used for our evaluation, while Table \ref{acceleratorparameters} provides the parameters used for assessing the overhead of the peripherals in our evaluated accelerators. We set each laser diode to emit an input optical power of 10 mW (10 dBm)  (Table \ref{Table:2}). The parameters for the multiplexer and splitter were sourced from \cite{holylight}.

\begin{table}[h!]
\caption{TPC size (\textit{N}) and TPC Count (\#) at 4-bit precision across various data rates for various accelerator architectures.}
\label{scalabilityandareaproportionate}\centering
\begin{tabular}{|c|cccccc|}
\hline
\multicolumn{1}{|l|}{} & \multicolumn{6}{c|}{\textbf{Datarate}}                                                                                                   \\ \hline
                       & \multicolumn{2}{c|}{\textbf{1 GS/s}}                        & \multicolumn{2}{c|}{\textbf{5 GS/s}}                        & \multicolumn{2}{c|}{\textbf{10 GS/s}}   \\ \hline
\textbf{TPC} &
  \multicolumn{1}{c|}{\textbf{N}} &
  \multicolumn{1}{c|}{\textbf{\#}} &
  \multicolumn{1}{c|}{\textbf{N}} &
  \multicolumn{1}{c|}{\textbf{\#}} &
  \multicolumn{1}{c|}{\textbf{N}} &
  \textbf{\#} \\ \hline
\textbf{SOIPhAR}           & \multicolumn{1}{c|}{22} & \multicolumn{1}{c|}{132} & \multicolumn{1}{c|}{15} & \multicolumn{1}{c|}{155} & \multicolumn{1}{c|}{13} & 162 \\ \hline
\textbf{SiNPhAR}           & \multicolumn{1}{c|}{47} & \multicolumn{1}{c|}{50} & \multicolumn{1}{c|}{28} & \multicolumn{1}{c|}{95} & \multicolumn{1}{c|}{22} & 116 \\ \hline
\end{tabular}
\label{Table:3}
\end{table}

\begin{table}[]
\begin{threeparttable}[b]
\caption{Accelerator Peripherals and TPC Parameters {\cite{cases2022}}.}
\label{acceleratorparameters}
\begin{tabular}{|c|ccc|}
\hline
                           & \multicolumn{1}{c|}{\textbf{Power(mW)}} & \multicolumn{1}{c|}{\textbf{Latency}} & \textbf{Area($mm^2$)} \\ \hline
\textbf{Reduction Network} & \multicolumn{1}{c|}{0.050}              & \multicolumn{1}{c|}{3.125ns}          & 3.00E-5            \\ \hline
\textbf{Activation Unit}   & \multicolumn{1}{c|}{0.52}               & \multicolumn{1}{c|}{0.78ns}           & 6.00E-5            \\ \hline
\textbf{IO Interface}      & \multicolumn{1}{c|}{140.18}             & \multicolumn{1}{c|}{0.78ns}           & 2.44E-2            \\ \hline
\textbf{Pooling Unit}      & \multicolumn{1}{c|}{0.4}                & \multicolumn{1}{c|}{3.125ns}          & 2.40E-4            \\ \hline
\textbf{eDRAM}             & \multicolumn{1}{c|}{41.1}               & \multicolumn{1}{c|}{1.56ns}           & 1.66E-1             \\ \hline
\textbf{Bus}               & \multicolumn{1}{c|}{7}                  & \multicolumn{1}{c|}{5 cycles}         & 9.00E-3               \\ \hline
\textbf{Router}            & \multicolumn{1}{c|}{42}                 & \multicolumn{1}{c|}{2 cycles}         & 1.50E-2              \\ \hline
\textbf{DAC \cite{dac} }      & \multicolumn{1}{c|}{12.5}         & \multicolumn{1}{c|}{0.78ns}             & 2.50E-3              \\ \hline
\textbf{ADC(1 GS/s) \cite{adc1gbps} }      & \multicolumn{1}{c|}{2.55}         & \multicolumn{1}{c|}{0.78ns}             & 2E-3              \\ \hline
\textbf{ADC(5 GS/s) \cite{adc3gbps} }      & \multicolumn{1}{c|}{11}         & \multicolumn{1}{c|}{0.78ns}             & 21E-3              \\ \hline
\textbf{ADC(10 GS/s) \cite{adc5gbps} }      & \multicolumn{1}{c|}{30}         & \multicolumn{1}{c|}{0.78ns}             & 103E-3           \\ \hline
\textbf{EO MRM Operation}         &  \multicolumn{1}{c|}{1.4 pJ/bit}          & \multicolumn{1}{c|}{-}             &0.95E-4                    \\\hline
% \textbf{EO Tuning}         & \multicolumn{1}{c|}{80 $\mu$W/FSR}          & \multicolumn{1}{c|}{20ns}             & -                  \\\hline
% \textbf{TO Tuning}         & \multicolumn{1}{c|}{275 mW/FSR}         & \multicolumn{1}{c|}{4$\mu$s}             & -                  \\ \hline

\end{tabular}
% \begin{tablenotes}
%        \item $^1$All accelerators operate DAC at 1 GS/s.
%        \item $^2$HEANA-OS operate DAC at 10 GS/s.

% \end{tablenotes}
  \end{threeparttable}
\end{table}

\subsection{System-Level Evaluation Results}
In Fig. \ref{fpsandenergyefficieny}(a), the Normalized FPS results for various accelerators with a batch size of 1, operating at different datarates are presented. These results are normalized to SOIPhAR for ResNet50 \cite{resnet} at a datarate of 10 GS/s. SiNPhAR accelerators outperforms SOIPhAR accelerators in terms of gmean across four CNN models at all datarates. Specifically, at 1 GS/s, SiNPhAR achieves up to 1.7$\times$ better FPS than SOIPhAR. As the datarate increases to 5 GS/s, SiNPhAR exhibits further improvements in FPS over SOIPhAR, achieving up to 1.8$\times$ better FPS than SOIPhAR. These remarkable throughput improvements in SiNPhAR are attributed to two main factors. Firstly, the SiNPhAR architecture utilizes SiN-based active and passive devices to implement analog GeMM functions. The low optical signal losses in the SiNPhAR architecture, owing to the low-index contrast and absence of two-photon absorption (TPA) in SiN material, enable the support of a larger TPC size (\textit{N=47}) compared to that of SOIPhAR (\textit{N=22}). This larger TPC size, as shown in Table \ref{Table:3}, increases the size of the dot product operation \textit{N} and the number of parallel dot product operations \textit{M}, thereby enhancing overall throughput via improved parallelism. Secondly, a larger \textit{N} results in fewer buffer accesses of weight and input values, reducing the buffer access latency. This reduction in access latency improves FPS. Furthermore, as the datarate increases, the FPS of each accelerator decreases. At 5 GS/s and 10 GS/s, the \textit{N} value decreases for all accelerators, as indicated in Table \ref{Table:3}, leading to low parallelism and increased buffer accesses. This increase in access latency with higher datarates results in lower FPS for the accelerators.

\begin{figure}[h!]
  \centering
  \includegraphics[width=230pt]{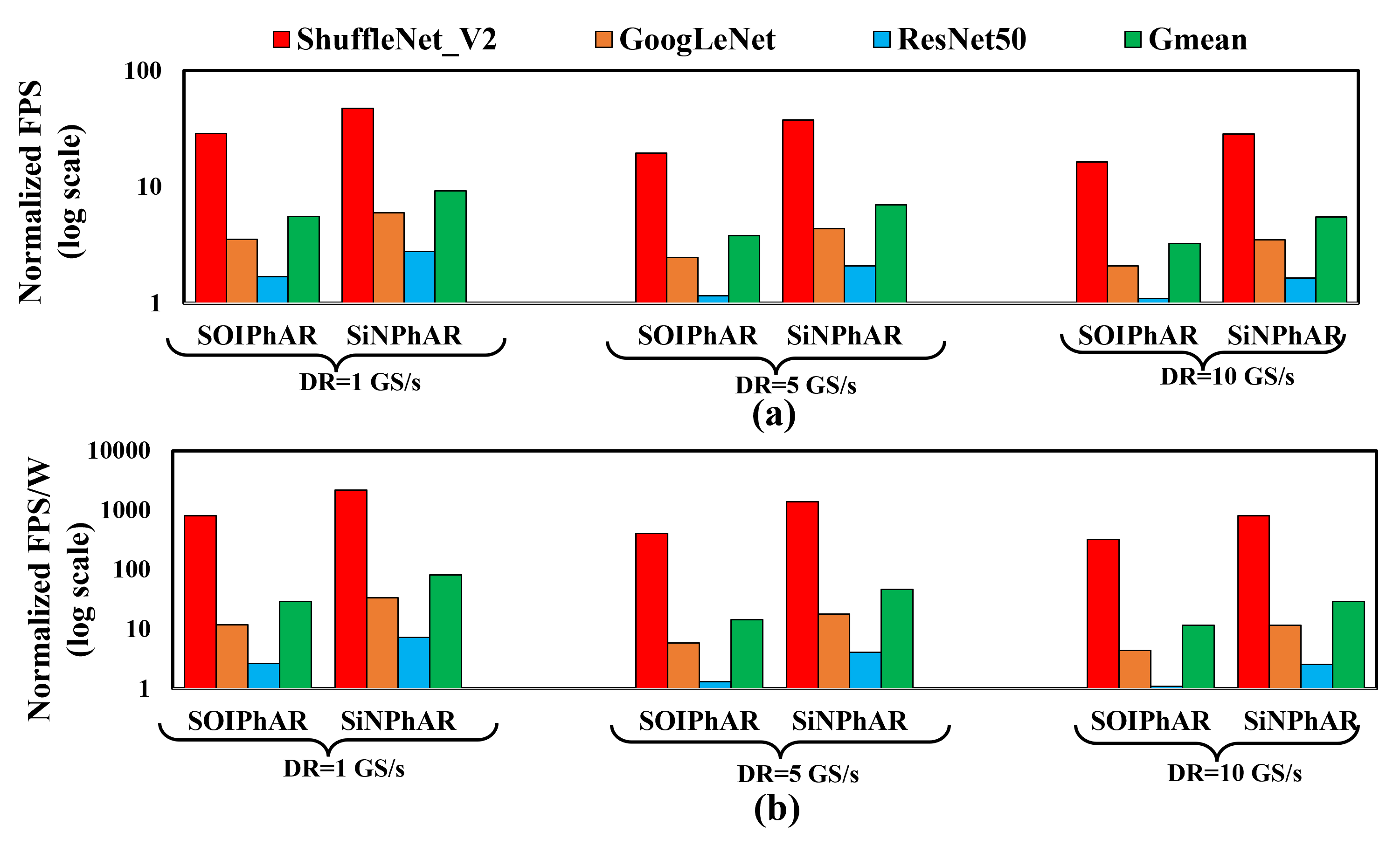}
  \caption{(a) Normalized FPS (log scale) (b) Normalized FPS/W (log scale) for SiNPhAR versus SOIPhAR accelerators with input batch size=1. Results of FPS and FPS/W are normalized w.r.t. SOIPhAR ResNet50 at 10 GS/s.} 
  \label{fpsandenergyefficieny}
\end{figure}

In Fig. \ref{fpsandenergyefficieny}(b), the energy efficiency (FPS/W) results are presented on a log scale for both SOIPhAR and SiNPhAR accelerators, using a batch size of 1 at various datarates. These results are normalized to SOIPhAR for ResNet50 at the datarate of 10 GS/s. Notably, the SiNPhAR accelerators demonstrate superior energy efficiency compared to the SOIPhAR accelerators. Specifically, at 1 GS/s, SiNPhAR achieves 2.8$\times$ better FPS/W compared to SOIPhAR, based on the Gmean across the CNNs. As the datarate increases to 5 GS/s, SiNPhAR achieves even better improvement over SOIPhAR, with 3.19$\times$ better FPS/W when compared to SOIPhAR.

These energy efficiency advantages of SiNPhAR stem from several factors. First, the improved throughput and reduced energy consumption of buffer accesses contribute to enhanced energy efficiency. As discussed earlier, the higher \textit{N} value supported by SiNPhAR results in improved parallelism, which, in turn, reduces dynamic energy consumption while maintaining higher throughput. Additionally, SiNPhAR requires overall fewer buffer accesses of input and weight values, leading to energy savings by reducing the energy consumption corresponding to buffer accesses. As the datarate increases, the peripheral components of the accelerator, such as ADCs and DACs, consume more power (as indicated in Table \ref{acceleratorparameters}). This additional power consumption decreases the achieved FPS/W for both SOIPhAR and SiNPhAR accelerators.

\section{Conclusion}
In this paper, we presented a novel SiN-based photonic GEMM Accelerator called SiNPhAR. Our SiNPhAR accelerator employs SiN-on-SiO${_2}$ based waveguides ITO-enabled SiN-on-SiO$_{2}$-basedmicroring modulators (MRMs) as input and weight elements, to implement analog GEMM functions. The key advantages of SiNPhAR over traditional SOI-based photonic GEMM accelerators lie in the absence of Two-Photon Absorption (TPA) nonlinearity and the low index contrast of the SiN-on-SiO$_{2}$ devices. These features enable SiNPhAR to experience significantly low optical signal losses compared to traditional SOI-based photonic GEMM accelerators, substantially enhancing its parallelism, throughput, and energy efficiency. To validate these benefits of our SiNPhAR accelerator, we evaluated its achievable parallelism and performance and compared it with a traditional SOI-based GEMM accelerator from prior work. Our analysis reveals that SiNPhAR supports at least 1.5$\times$ more multipliers than the prior SOI-based photonic GEMM accelerator. Furthermore, from the system-level performance analysis, SiNPhAR demonstrates at least 1.7$\times$ better throughput (FPS) while consuming at least 2.8$\times$ better energy efficiency (FPS/W) compared to the prior SOI-based GEMM accelerator.

\section*{Acknowledgments}
We thank the anonymous reviewers whose valuable feedback helped us improve this paper. We would also like to acknowledge the National Science Foundation (NSF) as this research was supported by NSF under grant CNS-2139167.

%In this paper, we presented the better performance of the SiNPhArch accelerator compared to the SiOPhArch accelerator across various datarates, across data rates and batch sizes. SiNPhArch consistently outperforms SiOPhArch in terms of normalized Frames Per Second (FPS), achieving up to 1.7× better performance at 1 GS/s and exhibiting further improvements at higher data rates. This remarkable throughput improvement is attributed to the utilization of SiN-based active and passive devices, which enable a larger TPC size and enhanced parallelism. Additionally, SiNPhArch demonstrates significantly higher energy efficiency, with up to 3.19× better FPS/W compared to SiOPhArch, owing to reduced energy consumption in partial sum reductions and fewer psums. While higher data rates lead to a decrease in FPS for all accelerators, SiNPhArch maintains its performance advantage.

\bibliographystyle{IEEEtran}
\bibliography{references}

\begin{thebibliography}{10}
\providecommand{\url}[1]{#1}
\csname url@samestyle\endcsname
\providecommand{\newblock}{\relax}
\providecommand{\bibinfo}[2]{#2}
\providecommand{\BIBentrySTDinterwordspacing}{\spaceskip=0pt\relax}
\providecommand{\BIBentryALTinterwordstretchfactor}{4}
\providecommand{\BIBentryALTinterwordspacing}{\spaceskip=\fontdimen2\font plus
\BIBentryALTinterwordstretchfactor\fontdimen3\font minus \fontdimen4\font\relax}
\providecommand{\BIBforeignlanguage}[2]{{%
\expandafter\ifx\csname l@#1\endcsname\relax
\typeout{** WARNING: IEEEtran.bst: No hyphenation pattern has been}%
\typeout{** loaded for the language `#1'. Using the pattern for}%
\typeout{** the default language instead.}%
\else
\language=\csname l@#1\endcsname
\fi
#2}}
\providecommand{\BIBdecl}{\relax}
\BIBdecl

\bibitem{dnnapplications1}
Y.~LeCun \emph{et~al.}, ``Deep learning,'' \emph{Nature}, 2015.

\bibitem{dnnapplications2}
W.~Liu \emph{et~al.}, ``A survey of deep neural network architectures and their applications,'' \emph{Neurocomputing}, 2017.

\bibitem{demirkiran2021electro}
C.~{Demirkirans \em et al.}, ``An electro-photonic system for accelerating deep neural networks,'' \emph{arXiv preprint arXiv:2109.01126}, 2021.

\bibitem{Baischer2021}
L.~Baischer \emph{et~al.}, ``Learning on hardware: A tutorial on neural network accelerators and co-processors,'' 2021.

\bibitem{holylight}
W.~{Liu \em et al.}, ``Holylight: A nanophotonic accelerator for deep learning in data centers,'' in \emph{IEEE DATE}.\hskip 1em plus 0.5em minus 0.4em\relax IEEE, 2019.

\bibitem{squeezelight}
J.~Gu \emph{et~al.}, ``Squeezelight: Towards scalable optical neural networks with multi-operand ring resonators,'' in \emph{DATE}, 2021.

\bibitem{deapcnn}
V.~{Bangari \em et al.}, ``Digital electronics and analog photonics for convolutional neural networks (deap-cnns),'' \emph{IEEE JSTQE}, 2019.

\bibitem{karen2020proceeding}
Q.~Cheng \emph{et~al.}, ``Silicon photonics codesign for deep learning,'' \emph{Proceedings of the IEEE}, 2020.

\bibitem{amm}
L.~Yang \emph{et~al.}, ``On-chip optical matrix-vector multiplier,'' in \emph{Optics and Photonics for Information Processing}.\hskip 1em plus 0.5em minus 0.4em\relax {SPIE}, 2013.

\bibitem{sunny2021crosslight}
F.~{Sunny \em et al.}, ``Crosslight: A cross-layer optimized silicon photonic neural network accelerator,'' in \emph{IEEE/ACM DAC}.\hskip 1em plus 0.5em minus 0.4em\relax IEEE, 2021.

\bibitem{acceleratorssurvey}
L.~{De Marinis \em et al.}, ``Photonic neural networks: A survey,'' \emph{IEEE Access}, 2019.

\bibitem{baets2016silicon}
R.~{Baets \em et al.}, ``Silicon photonics: Silicon nitride versus silicon-on-insulator,'' in \emph{OFC}.\hskip 1em plus 0.5em minus 0.4em\relax Optica Publishing Group, 2016, pp. Th3J--1.

\bibitem{ophir2010demonstration}
N.~{Ophir \em et al.}, ``Demonstration of 1.28-tb/s transmission in next-generation nanowires for photonic networks-on-chip,'' in \emph{2010 23rd Annual Meeting of the IEEE Photonics Society}.\hskip 1em plus 0.5em minus 0.4em\relax IEEE, 2010, pp. 560--561.

\bibitem{borghi2021modeling}
M.~{Borghi \em et al.}, ``On the modeling of thermal and free carrier nonlinearities in silicon-on-insulator microring resonators,'' \emph{Optica OE}, vol.~29, no.~3, pp. 4363--4377, 2021.

\bibitem{al2022scaling}
M.~{Al-Qadasi \em et al.}, ``Scaling up silicon photonic-based accelerators: Challenges and opportunities,'' \emph{APL Photonics}, 2022.

\bibitem{shiflett2020pixel}
K.~{Shiflett \em et al.}, ``Pixel: Photonic neural network accelerator,'' \emph{IEEE HPCA}, 2020.

\bibitem{cases2022}
S.~S. {Vatsavai \em et al.}, ``Photonic reconfigurable accelerators for efficient inference of cnns with mixed-sized tensors,'' \emph{IEEE TCAD}, 2022.

\bibitem{vatsavai2023sconna}
S.~{Vatsavai \em et al.}, ``Sconna: A stochastic computing based optical accelerator for ultra-fast, energy-efficient inference of integer-quantized cnns,'' \emph{arXiv preprint arXiv:2302.07036}, 2023.

\bibitem{padmaraju2014intermodulation}
K.~{Padmaraju \em et al.}, ``Intermodulation crosstalk characteristics of wdm silicon microring modulators,'' \emph{IEEE PTL}, 2014.

\bibitem{karempudi2023analysis}
V.~S.~P. Karempudi \emph{et~al.}, ``An analysis of various design pathways towards multi-terabit photonic on-interposer interconnects,'' \emph{JETCS}, dec 2023.

\bibitem{lee2008ultrahigh}
B.~G. {Lee \em et al.}, ``Ultrahigh-bandwidth silicon photonic nanowire waveguides for on-chip networks,'' \emph{IEEE PTL}, 2008.

\bibitem{blumenthal2018silicon}
D.~{Blumenthal \em et al.}, ``Silicon nitride in silicon photonics,'' \emph{Proceedings of the IEEE}, 2018.

\bibitem{ilie2022thermo}
S.~{Ilie \em et al.}, ``Thermo-optic tuning of silicon nitride microring resonators with low loss non-volatile sb 2 s 3 phase change material,'' \emph{Scientific Reports}, 2022.

\bibitem{chrostowski2014}
L.~{Chrostowski \em et al.}, ``Design methodologies for silicon photonic integrated circuits,'' \emph{SPIE SPOIC XVI}, 2014.

\bibitem{ma2015indium}
Z.~{Ma \em et al.}, ``Indium-tin-oxide for high-performance electro-optic modulation,'' \emph{Nanophotonics}, 2015.

\bibitem{li2019silicon}
E.~{Li}\em~et al., ``Silicon microring modulator with transparent conductive oxide gate,'' \emph{IEEE OI}, pp. 1--2, 2019.

\bibitem{alexandermit2022}
A.~{Sludds \em et al.}, ``Delocalized photonic deep learning on the internet’s edge,'' \emph{Science}, 2022.

\bibitem{BhaskaranPCA2022}
F.~Brückerhoff-Plückelmann \emph{et~al.}, ``A large scale photonic matrix processor enabled by charge accumulation,'' \emph{Nanophotonics}, 2022.

\bibitem{Wang2021-ISCAS}
C.-C. Wang \emph{et~al.}, ``67.5-fj per access 1-kb sram using 40-nm logic cmos process,'' in \emph{ISCAS}, 2021.

\bibitem{shufflenet}
X.~{Zhang \em et al.}, ``Shufflenet: An extremely efficient convolutional neural network for mobile devices,'' 06 2018, pp. 6848--6856.

\bibitem{googlenet}
C.~{Szegedy \em et al.}, ``Going deeper with convolutions,'' in \emph{CVPR}, 2015.

\bibitem{resnet}
K.~{He \em et al.}, ``Deep residual learning for image recognition,'' 06 2016, pp. 770--778.

\bibitem{dac}
F.~N.~U. Juanda \emph{et~al.}, ``A 10-gs/s 4-bit single-core digital-to-analog converter for cognitive ultrawidebands,'' \emph{TCS}, 2017.

\bibitem{adc1gbps}
D.-R. Oh \emph{et~al.}, ``An 8b 1gs/s 2.55mw sar-flash adc with complementary dynamic amplifiers,'' in \emph{IVLSIC}, 2020.

\bibitem{adc3gbps}
Y.-S. Shu, ``A 6b 3gs/s 11mw fully dynamic flash adc in 40nm cmos with reduced number of comparators,'' in \emph{VLSIC}, 2012.

\bibitem{adc5gbps}
M.~Guo \emph{et~al.}, ``A 29mw 5gs/s time-interleaved sar adc achieving 48.5db sndr with fully-digital timing-skew calibration based on digital-mixing,'' in \emph{VLSIC}, 2019.

\end{thebibliography}

\end{document}